\documentclass[11pt,english]{article}
\usepackage[T1]{fontenc}
\usepackage[english]{babel}

\usepackage[top=1in,bottom=1in,left=0.75in,right=0.75in]{geometry}
\usepackage{times}
\usepackage{amsmath,amssymb,amsfonts}
\usepackage{graphicx}
\usepackage[super]{nth}

\usepackage{natbib}
\usepackage{booktabs}
\usepackage{caption}
\usepackage{subcaption}
\usepackage{float}

\usepackage{hyperref}
\hypersetup{
  colorlinks   = true,
  linkcolor    = blue,
  citecolor    = blue,
  urlcolor     = cyan
}

\usepackage{titlesec}
\titleformat{\section}
  {\normalfont\large\bfseries}{\thesection.}{1em}{}
\titleformat{\subsection}
  {\normalfont\normalsize\bfseries}{\thesubsection}{1em}{}
\titlespacing*{\section}{0pt}{12pt}{6pt}
\titlespacing*{\subsection}{0pt}{12pt}{6pt}

\usepackage[flushmargin]{footmisc}

\usepackage[stretch=10,shrink=10,step=1]{microtype}
\usepackage{enumitem}

\usepackage{tabularx}
\newcolumntype{L}{>{\raggedright\arraybackslash}X}
\begin{document}
\twocolumn[
  \begin{center}
    {\LARGE Comparing Powerwise (PWR) and the NCAA Power Index (NPI): \\ Advising the NCAA Men's Division I Lacrosse Committee\par}
    \vskip 1.5em
    {\large
      \lineskip 2em
      \begin{tabular}[t]{c}
        Lawrence Feldman$^{1}$ \\
        \texttt{laf@laxnumbers.com}
        \and
        Matthew Bomparola$^{2}$ \\
        \texttt{matthewbomparola@gmail.com}
      \end{tabular}\par}
    \vskip 1.5em
  \end{center}
  
  \vskip 0.5em
  \noindent\textbf{Abstract:} This memo compares two methods—Powerwise (PWR) and the NCAA Power Index (NPI)—that aim to rank NCAA Division I, II, and III teams on the basis of deservedness of an invite to end-of-season championship tournaments. It finds that while the NPI might be a fit for sports like hockey, it falls short of the PWR method for use in ranking team sports that regularly feature somewhat wider margins of victory, including football, basketball, and lacrosse. In comparing the methods, this memo highlights differences in i) accuracy; ii) procedural integrity; iii) objectivity; iv) reproducibility; v) simplicity; and vi) stability; before drawing conclusions.
  
  
  \vskip 1.5em
]

\footnotetext[1]{Lawrence Feldman played lacrosse at the University of Pennsylvania, received a PhD from Auburn University and attended Stanford University as a Post Doc. He worked as an engineer/computer analyst in the USAF and worked in scientific laboratories, including Lawrence Livermore Laboratory (LLL) and Los Alamos National Laboratory (LANL). In addition, he worked for NASA and was a software architect for Intel Corp. Feldman has served on the Coaches Committee for Division I men's lacrosse for the past two years as an advisor. He founded Laxpower, a web site that rated lacrosse teams from 1997 to 2019.}

\footnotetext[2]{Matthew Bomparola has played tennis for two decades and has worked as a coach and professional hitting partner. He received a degree from Princeton University in '21 and works as a data analyst, writer, and journalist.}

\section{Introduction}
Powerwise (PWR) was designed in conversation with the NCAA Lacrosse Selection and Ranking Committee (SCR) to produce an objective ranking of teams by focusing on on-field performance in head-to-head and common opponent matchups and the goal difference between the winning and losing team for all games between all teams in a division during a given regular season.

The NCAA Power Index (NPI), on the other hand, was designed for use in DII and DIII sports and is driven primarily by win-loss records and adjustments, including a strength of schedule variable, quality win bonuses, and "dials" controlled by relevant committees.\footnote{While arguably more objective than the myriad systems previously used by Division II and III committees to determine tournament eligibility, the NPI's dials and its flawed strength of schedule component leave much to be desired; further discussion follows in the section on procedural integrity.} For further reference, the NPI is described in Appendix A and PWR is described in Appendix B.\footnote{Powerwise is described in detail in a whitepaper published by \cite{FeldmanBomparola2025}.}

This memo suggests that ranking methods for college sports that rely on win-loss records (like the NPI) fall short of common-sense approaches based on margins of victory (like PWR) and the key indicators of deservedness of an at-large pick: performance on the field in head-to-head and common opponent matchups.

Those interested in the quantitative differences between margin of victory and win-loss ranking methods might read: a \cite{Barrow2013a} analysis that comprehensively compares different ranking methods,\footnote{"For NCAAF and NCAAB datasets, the Nemenyi test concludes that the implementations utilizing score-differential data are usually more predictive than those using only win-loss data". The Ratings Percentage Index (RPI) was among the win-loss methods analyzed by Barrow et al. and is an index with a methodology that resembles the NPI's approach of weighting measures of team performance and strength of schedule.} an analysis by \cite{PaulWilson2015} on bias in rankings produced by the Ratings Percentage Index (RPI),\footnote{Paul and Wilson attribute bias to a lack of consideration for score data: "When using the computer rankings of Ratings Percentage Index (RPI), which only uses wins and losses in its analysis, bias in the selection process seems evident. No bias is revealed when the Sagarin "Predictor" rating is substituted... [which] uses margin of victory and results compared to expectations in its analysis..."} a report on a hybrid win-loss and score-differential method developed by \cite{Annis2005},\footnote{Annis and Craig suggest the use of a hybrid method as superior to win-loss alternatives.} comments by \cite{Chartier2011} on the Colley and Massey methods,\footnote{The "point differential parameter $\beta$ in the Massey model gives additional interpretability that is absent from the bias-free colley method, which considers only wins and losses."} and justification for including margins of victory in \cite{Massey1997}.\footnote{In Massey's method, "a team's rating is the mathematical sum of its average margin of victory and its average schedule strength... [going] beyond wins and points to determine how much respect a team really deserves for its performance.}

The following chapters on accuracy, procedural integrity, objectivity, reproducibility, simplicity, and stability support the claim that PWR is better suited than NPI to rank team sports like lacrosse, football, and basketball that regularly feature significant margins of victory, especially when data are relatively sparse, as in NCAA Division I, II, and III regular-season competition.

\section{Analysis}

\subsection{Accuracy}
The accuracy of a ranking method is limited by the quality and quantity of its data. College sports feature datasets with very little redundancy (often one or two games per pair of teams per season), sparse distribution (few games linking conferences), and, naturally, external randomness.

The evidence presented in this section might seem "cherry-picked," but the purpose of this memo is to point out the differences between the NPI and PWR, and these cases are illustrative examples rather than conclusive proof. 

The data used in the case studies presented below are presented in Tables 1 and 2 and consist of real-world NPI and simulated PWR ratings for the 2025 NCAA Men's and Women's Division III Lacrosse regular seasons.

\begin{table}[!htbp]
\centering
\caption{Results of Men’s NPI and PWR \\(Men’s Division III Lacrosse, 2025)}
\label{tab:d3-npi-pwr-compact}
\small
\setlength{\tabcolsep}{1pt}
\renewcommand{\arraystretch}{1.06}
\begin{tabular}{lcc@{\hspace{0.4em}}lcc}
\toprule
Team & NPI & PWR & \multicolumn{1}{c}{\ldots} & NPI & PWR \\
\midrule
Tufts (18-0)       & 1  & 1              & Cortland (17-2)       & 11 & 9 \\
Salisbury (19-0)   & 2  & 2              & Amherst (10-5)        & 12 & 7 \\
RIT (18-1)         & 3  & 4              & Ohio Wes. (15-3)      & 13 & \textbf{30} \\
C.\ Newport (16-3) & 4  & 5              & Stevenson (14-5)      & 14 & 18 \\
Bowdoin (13-3)     & 5  & 3              & St.\ Law. (11-6)      & 15 & 13 \\
York (16-3)        & 6  & \textbf{19}    & Dickinson (13-5)      & 16 & 20 \\
Wesleyan (12-5)    & 7  & 6              & Denison (13-6)        & 17 & 27 \\
W.\ \& Lee (16-3)  & 8  & \textbf{21}    & Middlebury (9-8)      & 18 & 10 \\
Gettysburg (13-4)  & 9  & 12             & Lynchburg (13-6)      & 19 & 25 \\
RPI (14-3)         & 10 & 8              & St.\ J.\ Fish. (15-4) & 20 & 26 \\
\bottomrule
\end{tabular}
\end{table}

\subsubsection{Case I: Bubble Teams}
Consider Table 1—the top five teams in this season are ranked similarly, with only positions 3, 4, and 5 slightly shuffled. This consistency is typical, as rating systems tend to agree on the strongest and weakest teams in a given data set. However, significant discrepancies arise among the "bubble teams."\footnote{Bubble teams are teams on the cusp of tournament qualification, usually roughly between ranks 8 and 14.}

The NPI ranks York, Washington \& Lee, and Ohio Wesleyan \nth{6}, \nth{8}, and \nth{13}, respectively. Powerwise, on the other hand, does not consider any of the three teams tournament contenders, placing them at \nth{19}, \nth{21}, and \nth{30}. Which system is closer to the truth?

Examining York's regular-season games, a game played against Albright, an NPI-ranked \nth{210} team with a 1-16 record, stands out. York only managed to eke out a narrow win, 11-9, despite playing at home—a clear underperformance against a weak opponent. 

Since the NPI disregards game scores, no explicit or implicit penalty is applied to York for its underperformance. PWR's Power Ratings, on the other hand, consider the closeness of this game in determining measures of team strength, slightly lowering York's PR rating, thus affecting PWR pairwise matchups not decided by on-field results. While no one game is enough to drop a team's ranking out of the top 10, multiple underperformances plus overperformances by similarly-ranked competitors might.

On the topic of overperformance: the NPI fails to capture the competitiveness of Middlebury's—NPI rank \nth{18}, PWR rank \nth{10}—and St. Lawrence's—NPI rank \nth{15}, PWR rank \nth{13}—results against their premier conference (NESCAC and LIBERTY) opponents. Whereas neither team's win-loss record is particularly remarkable, a further examination of their margins of victory (and loss) shows that they were competitive against some of the very best teams in DIII, deserving a rank that reflects their performance in close losses against difficult opponents.

Case I, therefore, indicates that ratings based on game scores (like PWR) are likely \textbf{more accurate} than ratings based exclusively on wins and losses (like the NPI), all other factors being equal, as they better capture the nuance of wins against easy opponents and losses against difficult ones—indicators of strength and deservedness. 

\subsubsection{Case II: Predictive Power}

Next, focus on a game played between Washington \& Lee and Gettysburg in the round of 16 of the 2025 men's DIII championship tournament. 

Whereas the NPI ranks Washington \& Lee above Gettysburg—\nth{8} vs. \nth{9}, PWR correctly predicts that Gettysburg is actually the stronger team; and they go on to earn a 14-12 win at Washington \& Lee's home stadium during the championship tournament. 

As an isolated result, this prediction is unremarkable, but the same is true on a more dramatic scale for Cortland and York, as Powerwise correctly predicts—by ranking Cortland \nth{9} and York \nth{21}—Cortland's unexpected (from the NPI's point-of-view) victory at York's home stadium. 

Case II, therefore, while somewhat circumstantial, provides evidence that PWR's approach is somewhat \textbf{more accurate} at predicting team strength than the NPI.\footnote{Predicting results or team strength is not the aim of either method, but deservedness likely correlates.} Further case study and statistical analysis is welcome and left to future study.

\subsubsection{Case III: Strength of Schedule}

As a final case study, consider the 2025 Women's DIII Lacrosse season presented in Table 2. As expected, the first nine NPI and PWR ranks are relatively well-aligned, if shuffled. The two major disagreements between the NPI and PWR are Sewanee—which has an NPI ranking of \nth{11} and a PWR ranking of \nth{58}—and Western Connecticut—NPI of \nth{18} and PWR of \nth{54}. Which pair of ranks is more accurate, and why?

\begin{table}[!htbp]
\centering
\caption{Results of Women’s NPI and PWR \\(Women’s Division III Lacrosse, 2025)}
\label{tab:w-d3-npi-pwr-compact}
\small
\setlength{\tabcolsep}{1pt}
\renewcommand{\arraystretch}{1.06}
\begin{tabular}{lcc@{\hspace{0.3em}}lcc}
\toprule
Team & NPI & PWR & \multicolumn{1}{c}{\ldots} & NPI & PWR \\
\midrule
Middlebury (16-1)    & 1  & 1              & Sewanee (17-0)     & 11 & \textbf{58} \\
Tufts (17-1)         & 2  & 2              & Rowan (15-3)       & 12 & 18 \\
F. \& M. (17-2)      & 3  & 7              & Salisbury (14-4)   & 13 & 12 \\
Colby (14-3)         & 4  & 3              & TCNJ (14-4)        & 14 & 20 \\
Gettysburg (16-3)    & 5  & 8              & Pomona-P. (17-2)   & 15 & 22 \\
St. J. Fisher (18-0) & 6  & 4              & Colorado C. (18-3) & 16 & 25 \\
Wesleyan (13-5)      & 7  & 5              & Stevens (15-4)     & 17 & 16 \\
York (16-3)          & 8  & 10             & W. Conn. (16-2)    & 18 & \textbf{54} \\
Amherst (11-5)       & 9  & 6              & Haverford (14-5)   & 19 & 15 \\
W. \& Lee (16-4)     & 10 & 17             & Denison (15-4)     & 20 & 28 \\
\bottomrule
\end{tabular}
\end{table}

In the championship tournament, the two discrepancies were decisively beaten by their opponents, Sweanee by Washington \& Lee 19-5 and Western Connecticut by Rowan 20-5. Although not conclusive, these results are evidence that the NPI may have overrated both teams. A clue as to why lies in the quality of their regular season schedules. 

Although Sewanee earned an undefeated win-loss record in their regular season games, they played exclusively against weak teams. Western Connecticut played a somewhat-stronger schedule, but neither they nor Sewanee won their regular season games against these weak teams convincingly enough (with a large enough margin of victory) to justify their placement as top 20 teams.

Case III thus calls into question the strength of schedule calculation used by the NPI and, in particular, the dial it uses to determine the relative weight of a team's win-loss record and the strength of their opponents; dials that, while adjustable, fail to consistently strike an accurate balance.\footnote{To be later discussed in more detail.} 

\subsubsection{Conclusions on Accuracy}
Ranking methods for college sports suffer from a lack of accuracy due to sparse data points linking teams and conferences. These three example cases highlight a major flaw that further weakens the NPI's ranking methodology: relevant (important!) information is disregarded—and often discarded\footnote{Victories against weak opponents that would worsen a team's NPI rating are thrown out. PWR's goal differential approach implicitly ensures that decisive wins against weaker opponents do not worsen a team's Power Rating, eliminating the need to delete data.}—despite an already information-poor environment. 

Win-loss records are not as rich as goal differences for use in comparing team strengths, and more importantly, deservedness of a position in an end-of-season championship tournament. 

To sum it up: PWR often yields more accurate rankings, particularly for teams in the middle tier, and the ommission of goal data likely explains most of the discrepancies between the NPI and PWR systems.

\subsection{Procedural Integrity}
Procedures are the set of rules used by sports ranking systems to produce results that (ideally) correspond to desired aims—in this case, deservedness of an at-large bid to an end-of-season championship tournament. 

This section, therefore, discusses whether the NPI and PWR's results are consistent  with their respective procedures; and the following, on that note, discusses whether their results are consistent with their respective data inputs.\\

To begin, recall that the NPI has two components:\\
\newcounter{myroman}
\begin{list}{\Roman{myroman}.}{%
    \usecounter{myroman}%
    \setlength{\leftmargin}{1.5em}
    \setlength{\labelwidth}{1.5em}%
    \setlength{\labelsep}{0.5em}
    \setlength{\itemsep}{2pt}
    \setlength{\topsep}{0pt}
    \setlength{\parsep}{0pt}%
    \setlength{\partopsep}{0pt}%
}
  \item \textbf{Team Performance}, as determined by win-loss
  reco- rds, with no adjustment for the quality of the opponent. 
  \item \textbf{Strength of Schedule}, calculated as the average NPI of a team's opponents.\\
\end{list}

Notably, as far as the "team performance" term is concerned, a big win against a strong opponent counts the same as a small win against a weak one. To address this obvious issue, the NPI calculates a second term, the "strength of schedule," and multiplies the rating produced by each term by a variable called a "dial" that adds weight based on relative importance.\footnote{The NPI actually uses four adjustable dials: the one discussed herein that balances win-loss and strength of schedule ratings, a dial for the home-field advantage, a dial for bonus points awarded in big wins, and a dial used specifically in NCAA Hockey for overtime wins. These dials are intended to make the NPI adaptable to different sports and are adjustable once every two years.} 

Importantly, this dial cannot remain "fixed" for all divisions or year-to-year—as critics of an old index called the Ratings Percentage Index (RPI) might recognize, the relative importance of terms I and II naturally depends on the distribution of conference and schedule strength in a given division and year. 

Thus, it must vary subject to committee input—intro-
ducing subjectivity, a topic further discussed in the next section.\footnote{An issue with mathematical methods that include arbitrary variables is that they often present as "objective" despite making use of non-result inputs.} 

The procedural problem here is simple: in most, if not all cases, the NPI's dials will fall at least a little short of the most accurate balance between team performance and strength of schedule, requiring some outside adjustments. 

What measure of "deservedness," then, does the NPI serve? One that is procedurally codified (and therefore verifiable), or one that depends on a committee's perception  of how different teams or conferences perform during the regular season? It depends on the dial.\\

In contrast with the NPI, Powerwise integrates team performance and opponent strength seamlessly by solving a system of linear equations for all teams in a division, taking game scores into account, producing a Power Rating (PR) for each team. 

PRs, used only when head-to-head and common opponents results are inconclusive, balance the relative importance of a team's performance against a given opponent without the need for an adjustable dial; plus, they do so by considering the strength of a team's schedule in relation to all other teams in the division, rather than just in relation to the teams they play.

As a result, PWR's fixed mathematical procedures automatically value blow-out wins against good opponents as more important to, say, blow-out wins against weak opponents; and as much more important to close wins against weak opponents. The NPI needs outside help to do the same, albeit still imperfectly.

\subsubsection{Conclusions on Procedural Integrity}
Whereas the NPI might seem procedurally complete and self-contained despite requiring outside adjustment, PWR, as presently conceived, is more fully specified, fully reproducible from public inputs, and free of post-hoc committee tuning. 

\subsection{Objectivity}
A ranking method is objective when its results follow from fixed, public inputs rather than from discretionary adjustment. 

Objective, well-specified, and public measures of performance are desirable insofar as it is difficult to find a committee that is perfectly unbiased, well-informed, and incentivized to produce optimally-accurate rankings. Unfortunately, as discussed, it is difficult to strike the balance between team performance and schedule strength without introducing errors. 

In addition, even a perfect closed-door committee debating such tight decisions risks being perceived as a "byzantine black box," producing "speculation, second guessing, and debate," or even doubt, regarding tournament selection decisions. (\cite{Coleman2010,Colley2002})\\

As mentioned, the NPI requires outside subjective input to achieve a balanced solution. Powerwise does not. All inputs needed to produce a Powerwise ranking list for a given year are as follows:\\

\begin{list}{\Roman{myroman}.}{%
    \usecounter{myroman}%
    \setlength{\leftmargin}{1.5em}
    \setlength{\labelwidth}{1.5em}%
    \setlength{\labelsep}{0.5em}
    \setlength{\itemsep}{2pt}
    \setlength{\topsep}{0pt}
    \setlength{\parsep}{0pt}%
    \setlength{\partopsep}{0pt}%
}
  \item \textbf{Scores}, for every regular-season game.
  \item \textbf{Location}, home or away, where the games were played.\\
\end{list}

That's all—no adjustable dials, no secrets. Assuming the method (described in the Powerwise whitepaper published by \cite{FeldmanBomparola2025}) is sound, and that enough regular season games have been played, PWR thus produces an objective ranking of any set of teams, avoiding favoring teams with strong or weak schedules without the need for outside input.

The major benefits of such a system are its reproducibility and simplicity—to be discussed in the next sections—allowing for informed debate and, hopefully, widespread acceptance in the college sports community.

\subsection{Reproducibility}
Reproducibility—i.e., the ability of others to verify results generated by a given method—is essential for transparency and allows teams to pursue strategic planning. 

As discussed, Powerwise is based on simple, publicly-available data and can easily publish both the results of its pairwise comparisons and Power Ratings, enabling coaches and analysts to predict future ratings based on the outcomes of the remaining schedule.

The NPI, due to its opaque and ad hoc components, cannot offer the same level of reproducibility throughout the season or even during the tournament selection process.

\subsection{Simplicity or "Understandability"}
On a similar note, simplicity (or the perception thereof) is key to buying the trust of coaches, players, and fans. 

The NPI's "team performance" term is relatively simple, but its strength of schedule term and the "dials" used to value each of its components require justification and will likely spark debate. Still, its procedural methodology can be described simply enough, ignoring the ad hoc inclusions and its arbitrary components.\\ 

PWR is similarly simple and can be described in fewer than 40 words: it awards points to teams in pairwise matchups against all other teams using a three-tiered system that includes on-field performance in i) head-to-head matchups and ii) common-opponent games and iii) Power Ratings as a tie-breaker. 

A positive of PWR is that nearly 60\% of its pairwise points are entirely determined by on-field results. As a result, an analyst advising a team's coaching staff might clearly and simply explain the origins of a team and their opponent's PWR points—building trust, enabling verifiability, and assisting strategy.\\

Of course, the formulae used by both the NPI and PWR (Power Ratings) feature mathematics that require computerized calculations to "converge" on a solution. For the average fan, it's complicated, but linear algebra isn't rocket science, and, because PWR's rankings are based on publicly-available data, a somewhat-savvy programmer might easily copy the code.

\subsubsection{Conclusions: Simplicity}
PWR's system, if slightly more complicated at first glance, is simpler to justify and reproduce thanks to a bit of extra data (in the form of goal differences) and math (in the form of Power Ratings) that allow for the exclusion of ad hoc and arbitrary adjustments.

\subsection{Stability}
The NPI is computed iteratively, solving for team ratings with systems of equations until they converge on a stable solution.

Interestingly, the NPI formula is such that, to know a team's NPI, one must know its strength-of-schedule rating, which is, in turn, a function of its opponents' NPIs. In other words, the NPI variable appears on both sides of the equation, meaning that the system is non-linear (mathematically speaking).

Because of this, there are combinations of dials and results in which convergence becomes increasingly difficult. In fact, depending on the dial positions, the iteration has been seen to not converge, becoming unstable and unsolvable.

To the authors' knowledge, neither Powerwise nor its Power Ratings suffer from non-convergence.

\section{Conclusions}
The NPI was intended for use in all divisions of all NCAA team sports—an admirable but ambitious goal. Unfortunately, the sources, case studies, and procedural analysis in this memo show that the NPI's win-loss, RPI-like methodology is not as accurate, objective, or reproducible as Powerwise, at least for use in sports like lacrosse, basketball, and football that feature relatively few games and large goal differentials. \\

Powerwise is a superior method for ranking teams in terms of:\\

\begin{list}{\Roman{myroman}.}{%
    \usecounter{myroman}%
    \setlength{\leftmargin}{1.5em}
    \setlength{\labelwidth}{1.5em}%
    \setlength{\labelsep}{0.5em}
    \setlength{\itemsep}{2pt}
    \setlength{\topsep}{0pt}
    \setlength{\parsep}{0pt}%
    \setlength{\partopsep}{0pt}%
}
  \item \textbf{Accuracy}, which depends on data. The win-loss records used by the NPI are too information-poor for college sports rankings, especially when goal differentials exist.
  \item \textbf{Determining schedule strength}, which is a key component of deservedness of an at-large bid. Despite trying to measure schedule strength, the NPI improperly values a big win against a strong opponent as nearly equal to a small win against a weak one.
  \item \textbf{Procedures}, which should be common-sense, purp-
  ose-built, and unbiased. Systems that use dials to balance team performance and strength of  schedule are somewhat subjective, adjustable, and therefore less likely to consistently produce accurate rankings than systems without.
  \item \textbf{Objectivity}, which builds trust that a system does what it says it does, and fairly. Methods (like PWR) that use simple, public data are superior to methods that make use of arbitrary or subjective components (like the NPI), especially in terms of reproducibility and simplicity.
  \item \textbf{Reproducibility}, which is key to helping coaches, fans, and players analyze their chances and build strategies to achieve an at-large bid. Given a simple set of public data, a computer programmer can quickly reproduce PWR's results.
  \item \textbf{Simplicity}, which earns trust and fosters good sportsmanship. On-field results directly (no math) determine roughly 60\% of Powerwise points—the NPI's formulas, adjustments, and dials complicate understanding.
  \item \textbf{Stability}—the NPI's formula has been seen to fail to converge on a solution given certain combinations of dials and results. 
  \\
\end{list}

PWR's ability to directly incorporate real-world, head-to-head and common opponent outcomes gives it a significant edge over the NPI, which tends to oversimplify rankings and requires arbitrary adjustments. Finally, its transparency, simplicity, and reproducibility give it a better chance of widespread acceptance by the NCAA community.\\

All considered, Powerwise offers a more accurate, objective, and easier-to-understand system for certain sports that feature goal differentials, like NCAA Men's Division I, II, and III lacrosse, basketball, and football.

\onecolumn
\nocite{*}
\bibliographystyle{plainnat}
\bibliography{references}

\appendix

\newpage
\section*{Appendix A: NCAA Power Index (NPI) Explainer}
The NCAA Power Index (NPI) is a new procedure for selecting teams for NCAA championship tournaments. The old method was extensive and based on the Ratings Percentage Index (RPI). Despite being based on the same two components as the RPI—a team's performance (win-loss record) and its strength of schedule—the NPI's strength of schedule is much simpler, relying solely on the average of the opponents' NPIs.\\ 

The NPI is calculated as follows: 
\[
\mathrm{NPI}_{(i+1)}
=
\mathit{dial}\left(\frac{W}{W+L}\right)
+
(1-\mathit{dial})\sum_{i=1}^{n}\frac{\mathrm{NPI}_i}{n}
\]
where the NPI is the NCAA Power index; dial is a variable that determines the relative weight of performance and schedule strength (ranges from 0.0 to 1.0); W = wins; L = Losses; and n = the number of teams in the division.\\\\

\section*{Appendix B: Powerwise (PWR) Explainer}
Powerwise (PWR) pairs each team against every other team in the division in hypothetical matchups in which one team "wins" and earns a "Powerwise point." Per matchup, a team wins the Powerwise point if it has a better:\\ 

\begin{list}{\Roman{myroman}.}{%
    \usecounter{myroman}%
    \setlength{\leftmargin}{1.5em}
    \setlength{\labelwidth}{1.5em}%
    \setlength{\labelsep}{0.5em}
    \setlength{\itemsep}{2pt}
    \setlength{\topsep}{0pt}
    \setlength{\parsep}{0pt}%
    \setlength{\partopsep}{0pt}%
}
  \item \textbf{Head-to-head} record against the other team \\(continue to II. only in case of tie);
  \item \textbf{Common opponent} record against the other team \\(continue to III. only in case of tie);
  \item \textbf{Power Rating (PR)} than the other team.\\
\end{list}

All Powerwise points are tallied and the team with the highest number (i.e., the best record) is ranked \nth{1}, and the remaining teams are ordered accordingly. In men's DI lacrosse there are 77 teams, meaning that the maximum number of Powerwise points is 76, and the minimum 0. 

Power Ratings (PRs) are an iterative process in which the difference in PR between two teams is calculated to equal the hypothetical goal difference if those teams were to play on a neutral field. Home-field advantage is included and based on the average goals scored at home vs. away, but the HFA method of preference is left to the committees.\\ 

Power Ratings are calculated as follows: 
\[
\sum_{i=0}^{n} \sum_{j=0}^{m} \left( \mathrm{PR}_i - \mathrm{PR}_j \right)
=
\sum_{i=0}^{n} \sum_{j=0}^{m} \left( \mathrm{score}_i - \mathrm{score}_j \right) \pm \mathit{hfa}
\]
where every combination of teams i and j are considered, and hfa represents the home-field advantage.
\end{document}